\documentclass[aps,prl,reprint,tightenlines,superscriptaddress]{revtex4-2}

\usepackage[T1]{fontenc} 
\usepackage{chemformula} 
\usepackage{amsmath} 
\usepackage{makecell,tabularx}
\setcellgapes{3pt}
\usepackage{braket}
\usepackage{lmodern}

\usepackage[separate-uncertainty=true, 
            multi-part-units=single]{siunitx} 

\DeclareSIUnit{\sample}{S}
\DeclareSIUnit\bit{b}

\usepackage{xargs}
\usepackage[disable,colorinlistoftodos,prependcaption,textsize=tiny]{todonotes}
\newcommandx{\unsure}[2][1=]{\todo[linecolor=red,backgroundcolor=red!25,bordercolor=red,#1]{#2}}
\newcommandx{\change}[2][1=]{\todo[linecolor=blue,backgroundcolor=blue!25,bordercolor=blue,#1]{#2}}
\newcommandx{\FIXME}[1][1=]{\todo[linecolor=blue,backgroundcolor=blue!25,bordercolor=blue,#1]{FIXME}}
\newcommandx{\info}[2][1=]{\todo[linecolor=OliveGreen,backgroundcolor=OliveGreen!25,bordercolor=OliveGreen,#1]{#2}}
\newcommandx{\improvement}[2][1=]{\todo[linecolor=Plum,backgroundcolor=Plum!25,bordercolor=Plum,#1]{#2}}
\newcommandx{\addref}[1][1=]{\todo[linecolor=red,backgroundcolor=red!25,bordercolor=red,#1]{Add reference}}

\newcommand\gtwo[2][]{\ifthenelse{\isempty{#2}}{$g^{(2)}(0)$}{\ifthenelse{\isempty{#1}}{$g^{(2)}(0)=#2$}{$g^{(2)}(0)=#2\pm#1$}}}

\usepackage{amsmath,amsfonts,amssymb}
\usepackage{mathrsfs}
\usepackage{xparse}
\usepackage{physics}
\usepackage{booktabs}
\usepackage{graphicx}
\usepackage[utf8]{inputenc}
\usepackage[english]{babel}

\usepackage{graphicx}
\usepackage{multirow}%
\usepackage{mathrsfs}%
\usepackage[title]{appendix}%
\usepackage{xcolor}%
\usepackage{textcomp}%
\usepackage{manyfoot}%
\usepackage{booktabs}%
\usepackage{algorithm}%
\usepackage{algorithmicx}%
\usepackage{algpseudocode}%
\usepackage{listings}%
\usepackage[separate-uncertainty=true, 
            multi-part-units=single]{siunitx} 
\usepackage{makecell,tabularx}
\setcellgapes{3pt}
\usepackage{braket}
\setcitestyle{numbers,square}
\usepackage[T1]{fontenc}
\usepackage{hyperref}

\makeatletter
\providecommand\@reinserts{} 
\makeatother

\begin{document}
\title{Fourier State Tomography of Polarization-Encoded Qubits}
\author{Mohammed K. Alqedra}
 \thanks{These authors contributed equally to this work.}
 \affiliation{Department of Applied Physics, Royal Institute of Technology, Albanova University Centre, 10691, Stockholm, Sweden}

\author{Pierre Brosseau}
 \thanks{These authors contributed equally to this work.}
 \affiliation{Centre for Disruptive Photonic Technologies, TPI, Nanyang Technological University, 637371 Singapore, Singapore}
 \affiliation{Division of
Physics and Applied Physics, School of Physical and Mathematical Sciences, Nanyang Technological University,
637371 Singapore, Singapore}

\author{Ali W. Elshaari}
 \affiliation{Department of Applied Physics, Royal Institute of Technology, Albanova University Centre, 10691, Stockholm, Sweden}

\author{Jun Gao}
 \affiliation{Department of Applied Physics, Royal Institute of Technology, Albanova University Centre, 10691, Stockholm, Sweden}

\author{Anton N. Vetlugin}
 \affiliation{Centre for Disruptive Photonic Technologies, TPI, Nanyang Technological University, 637371 Singapore, Singapore}
 \affiliation{Division of
Physics and Applied Physics, School of Physical and Mathematical Sciences, Nanyang Technological University,
637371 Singapore, Singapore}

\author{Cesare Soci}
 \affiliation{Centre for Disruptive Photonic Technologies, TPI, Nanyang Technological University, 637371 Singapore, Singapore}
 \affiliation{Division of
Physics and Applied Physics, School of Physical and Mathematical Sciences, Nanyang Technological University,
637371 Singapore, Singapore}

\author{Val Zwiller}
 \email{zwiller@kth.se}
 \affiliation{Department of Applied Physics, Royal Institute of Technology, Albanova University Centre, 10691, Stockholm, Sweden}


\date{\today}

\begin{abstract}
Quantum state tomography is a central technique for the characterization and verification of quantum systems. Standard tomography is widely used for low-dimensional systems, but for larger systems, it becomes impractical due to the exponential scaling of experimental complexity with the number of qubits. Here, we present an experimental realization of Fourier-transform quantum state tomography for polarization-encoded photonic states. We validate the technique using weak coherent states and entangled photon pairs generated by a quantum dot and spontaneous parametric down-conversion source in the telecom wavelength. The reconstructed density matrices show excellent agreement with those obtained through conventional projective tomography, with calculated metrics such as fidelity and concurrence matching within error bars, confirming the reliability and accuracy of the technique. 
Fourier state tomography employs only a single rotating waveplate per qubit, thereby avoiding repeated adjustments across multiple waveplates and ensuring that the number of physical measurement settings scales linearly with the number of qubits, despite the exponential growth of the underlying state space. This reduction in optical configurations simplifies experimental overhead, making Fourier state tomography a practical alternative for multi-qubit characterization.

\end{abstract}

\maketitle

Quantum state tomography (QST) is a cornerstone of quantum information science that enables the complete reconstruction of an unknown quantum state from measured data \cite{James2001Oct, Hradil1997Mar, Matteo2004}. In photonic quantum systems, QST is particularly important for characterizing polarization-encoded qubits \cite{James2001Oct}, which are widely used in quantum communication \cite{Kimble2008Jun, Wehner2018Oct}, cryptography \cite{Bennett2014Dec, Bennett1992Feb, Gisin2002Mar}, and photonic quantum computing \cite{Knill2001Jan, OBrien2009Dec}. By measuring a set of complementary observables and applying appropriate reconstruction algorithms \cite{Hradil2004Aug}, QST enables a comprehensive understanding of the underlying quantum system.

Standard projective tomography utilizes a set of waveplates and polarizers to measure discrete projection bases, with each basis measurement requiring a distinct experimental configuration \cite{James2001Oct, ALTEPETER2005105}. Although this approach is practical for low-dimensional systems, such as polarization-entangled photon pairs, the experimental complexity\textemdash such as the number of measurements, configurations, and computational processing, grows exponentially with the number of qubits \cite{Gross2010Oct, Cramer2010Dec}. This exponential scaling presents a fundamental challenge, rendering projective tomography impractical for high-dimensional quantum systems. Incomplete measurement techniques, such as entanglement witnesses \cite{Bourennane2004Feb, Lu2007Feb, Guhne2009Apr, Huang2011Nov, Yao2012Apr, Wang2016Nov, Malik2016Apr, Chen2017Jan, Zhong2018Dec}, have emerged as an alternative approach to gain partial information about higher-dimensional quantum states while avoiding the exponential burden of full state tomography. Nevertheless, entanglement witnesses are typically tailored  for specific types of states or entanglement structures. Besides, they do not provide comprehensive information about the quantum state, thereby restricting their general applicability. While other resource-efficient methods, such as compressed sensing \cite{Gross2010Oct, Cramer2010Dec}, have been proposed, they often require specific assumptions about the state or prior knowledge, limiting their general applicability.
Fourier-transform quantum state tomography (FT-QST) has been proposed as a resource-efficient alternative, reducing the number of physical measurement configurations by employing only one single waveplate and polarizer per qubit mode \cite{Mohammadi2013Jan}. Instead of discrete measurements, FT-QST measures a pseudocontinuous, time-dependent signal as each waveplate rotates at a distinct frequency, enabling full state reconstruction via Fourier analysis. Because each qubit needs only a single rotating waveplate, the hardware/configuration overhead scales linearly with the number of qubits, in stark contrast to the exponential number of settings required by standard projective tomography. It should be noted that we still need to collect enough time samples (per Nyquist’s criterion) to capture all frequency components, but this pseudocontinuous scan avoids physically reconfiguring the apparatus in an exponential number of ways. Consequently, FT-QST is particularly advantageous for characterizing multi-qubit systems.
\begin{figure*}[htb]
\includegraphics[width=2\columnwidth]{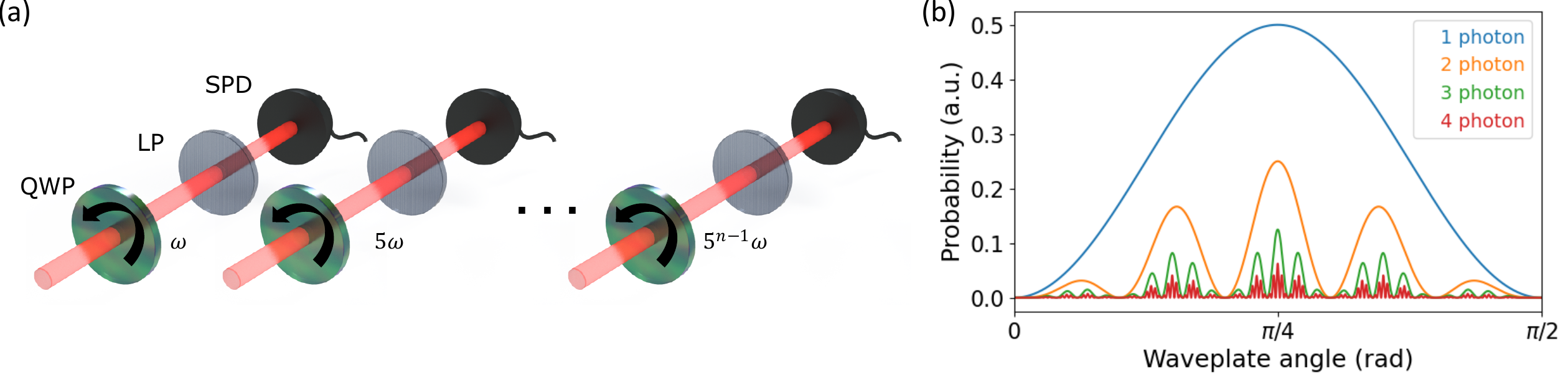}
\caption{\label{fig:setup_schematics}
(a) Fourier tomography of $n$-qubit photonic state. For every mode $n$, a quarter waveplate (QWP) rotating at $5^{n-1}\omega$ is used. A fixed linear polarizer (LP) is placed in front of every single photon detector (SPD). (b) Simulated probabilities as a function of waveplate rotation angle for one (blue), two (yellow) three (green) and four (red) photons with vertical polarization.
}
\end{figure*} 

In this work, we present the first experimental realization of FT-QST, demonstrating its feasibility for reconstructing quantum states of single- and two-qubit polarization states. The use of superconducting nanowire single-photon detectors (SNSPDs) is an additional advantage, as they offer higher detection efficiency with lower dark counts than semiconductor-based single-photon detectors. We first validate the technique by measuring polarization-filtered weak coherent states produced by a laser attenuated to the single-photon level. Subsequently, we extended the application of FT-QST to measure entangled photon pairs generated by an InAs/GaAs quantum dot in the telecom C-band \cite{Zeuner2021Aug}. We cross-verify the results with standard projective tomography and observe excellent agreement on key metrics such as fidelity, concurrence, and fine-structure splitting (FSS). 
We further validate FT-QST using polarization-entangled pairs generated by spontaneous parametric down-conversion (SPDC).
Throughout these experiments, we utilized a maximum-likelihood estimation algorithm to ensure the physicality of the reconstructed density matrices from the Fourier analysis of the measured data. Our findings demonstrate the reliability and accuracy of FT-QST for characterizing both single- and multi-qubit polarization states, offering a resource-efficient alternative to conventional methods.


The schematic of the setup for performing Fourier state tomography for an $n$-qubit photonic state $\rho$ is shown in Figure \ref{fig:setup_schematics}.a. It consists of $n$ modules, each measuring a qubit mode ,$m$, using a rotating quarter waveplate (angular frequency $\omega_m$), a fixed horizontal polarizer, and a single photon detector. 
We will briefly summarize the essential theory here, while referring the reader to Ref. \cite{Mohammadi2013Jan} for a more detailed derivation. The measurement probability of detecting photons in the horizontal polarization configuration for each mode, $p_n(t)$, is given by:
\begin{equation}
    p_n(t) = \frac{1}{2^n} \sum_{i_1,\dots,i_n=0}^{3} 
    S_{i_1,\dots,i_n} \prod_{m=1}^{n} \chi_{m,i_m}(t),
\end{equation}
where $S_{i_1,\dots,i_n}$ are the generalized Stokes parameters, and $\chi_{m,i_m}(t)$ is a time-dependent function that encodes the effect of the rotating waveplate in mode $m$ for a particular Pauli operator index $i_m$.
Due to the oscillatory phase shifts induced by the waveplate rotation, $\chi_{m,i}(t)$ can be written in terms of sines and cosines at integer multiples of $\omega_m$. Consequently, the product $\prod_{m=1}^{n} \chi_{m,i_m}(t)$ can be expressed as a finite sum of sines and cosines at the combined frequencies of $\{\omega_1 \dots \omega_m\}$. Consequently, $p_n(t)$ takes a Fourier-series expansion of the form:
\begin{equation}
p_n(t) = a_0 + \sum_{f \in \mathcal{F}}
\left[ a_f \cos(\Omega_f t) + b_f \sin(\Omega_f t) \right],
\end{equation}
where $\mathcal{F}$ is the set of all relevant frequency components arising from combinations of ${\omega_1, \dots, \omega_n}$. $\Omega_f$ represents a specific linear combination of the fundamental rotation frequencies. $a_f$ and $b_f$ are Fourier coefficients, which can be expressed as linear combinations of the Stokes parameters $S_{i_1,\dots,i_n}$. 
For the explicit form of $\chi_{m,i}(t)$, see Ref. \cite{Mohammadi2013Jan}. In the supplementary material, we present the special cases for a single- and two- qubit modes.

Here, we choose an integer ratio of 5 between successive waveplate rotation frequencies, as it cleanly eliminates degeneracies among the measured harmonic components and thereby enables unambiguous extraction of all generalized Stokes parameters \cite{Mohammadi2013Jan}. Smaller integer ratios either produce overlapping harmonics or fail to generate enough distinct frequency components for a full reconstruction (potentially leading to aliasing issues), making 5 the minimal ratio that ensures a complete, nondegenerate measurement. Choosing a larger ratio introduces additional frequency components that increase measurement overhead without providing any further advantage.
Figure \ref{fig:setup_schematics}.b shows a simulation of the measured signal for one (blue), two (yellow), three (green), and four (red) qubit modes when all photons are prepared in the vertical polarization $\ket{V}$. The oscillations increasingly combine multiple rotation frequencies and their harmonics as the number of qubits grows. This emphasizes how Fourier state tomography scales linearly in the number of waveplates—each rotating at a unique frequency—but gives rise to a richer frequency-domain signal whose harmonic content enables reconstruction of the higher-dimensional quantum state.
\begin{figure*}[htb]
\includegraphics[width=2\columnwidth]{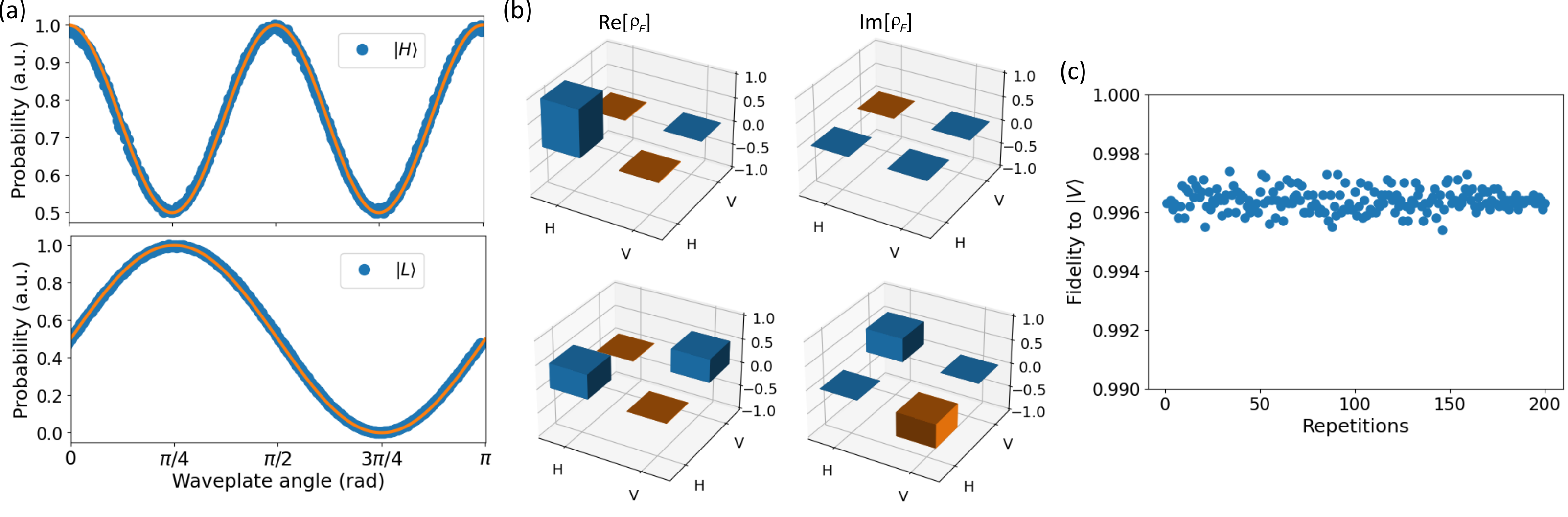}
\caption{\label{fig:single_photon_Polarimetry}
\textbf{Fourier tomography of single photon states.} (a) Measured (blue dots) and theoretical (orange line) signals for $\ket{H}$ (top) and $\ket{L}$ (bottom) polarization states. (b) Real and imaginary parts of the reconstructed density matrix for $\ket{H}$ (top) and $\ket{L}$ (bottom) polarization states from (a). Positive values are shown in blue, while negative values are represented in orange. (c) Repeated measurements of fidelity for $\ket{V}$ polarization state with a mean value of 0.996 and a standard deviation of less than 0.001 }
\end{figure*} 
To reconstruct a physical density matrix $\rho$ from the measured data, we use a maximum-likelihood estimation algorithm \cite{Hradil2004Aug}, ensuring that the reconstructed $\rho$ is positive semi-definite, Hermitian, and has unit trace.  First, we parametrize $\rho$ as $\rho = T \, T^\dagger$, where $T$ is a lower-triangular matrix with complex entries. Given a candidate $\rho$, we compute the generalized Stokes parameters $S_{i_1,\dots,i_n}$, which in turn map directly to the Fourier coefficients $\{a_f,b_f\}$. These coefficients define the fitted probabilities, $p_{\mathrm{fit}}(t)$, which are compared to the experimentally measured probabilities, $p_{\mathrm{measured}}(t)$, by minimizing the cost function:
\begin{equation}
    \label{eq:costfunction}
    C \;=\; \sum_{t}
    \Bigl[
      p_{\mathrm{measured}}(t) \;-\; p_{\mathrm{fit}}(t)
    \Bigr]^2,
\end{equation}
where the sum runs over all time bins or waveplate angle.
We use a gradient-based optimizer (e.g., L-BFGS-B) to perform this minimization. After convergence, we obtain $\rho$ from the optimized $T$ and normalize it so that $\mathrm{Tr}(\rho) = 1$. The resulting physical density matrix closely reproduces the observed data, enabling the calculation of fidelity, concurrence, and other figures of merit.

To estimate uncertainties in the reconstructed density matrices and derived the errors in figures of merit (e.g., fidelity and concurrence), we perform a Monte Carlo simulation of noisy measurement data \cite{James2001Oct, ALTEPETER2005105}. Specifically, we generate multiple synthetic datasets by sampling each measured probability from a normal distribution centered on the experimentally obtained value, with a standard deviation equal to its square root to approximate Poisson-like counting fluctuations. Each dataset is processed through the same maximum-likelihood reconstruction procedure, yielding a set of density matrices. We compute the mean and standard deviation of these matrices element-wise, as well as the spread in derived quantities such as fidelity and concurrence. This approach provides a robust estimate of the reconstruction’s sensitivity to statistical fluctuations and quantifies the uncertainties in both the density matrix elements and related physical observables.

\section*{Single-photon polarimetry}
To validate the FT-QST apparatus and the reconstruction algorithm, we first applied the technique in a single-photon polarimeter configuration. We used weak coherent polarization states generated by a laser attenuated to the single-photon level \cite{Kok2007Jan}. We verify the polarization state at classical light level using a commercial polarimeter (Thorlabs PAX1000IR2). These states provide a reliable approximation of single-photon states for calibration and testing.  
\begin{figure*}[htb]
\includegraphics[width=2\columnwidth]{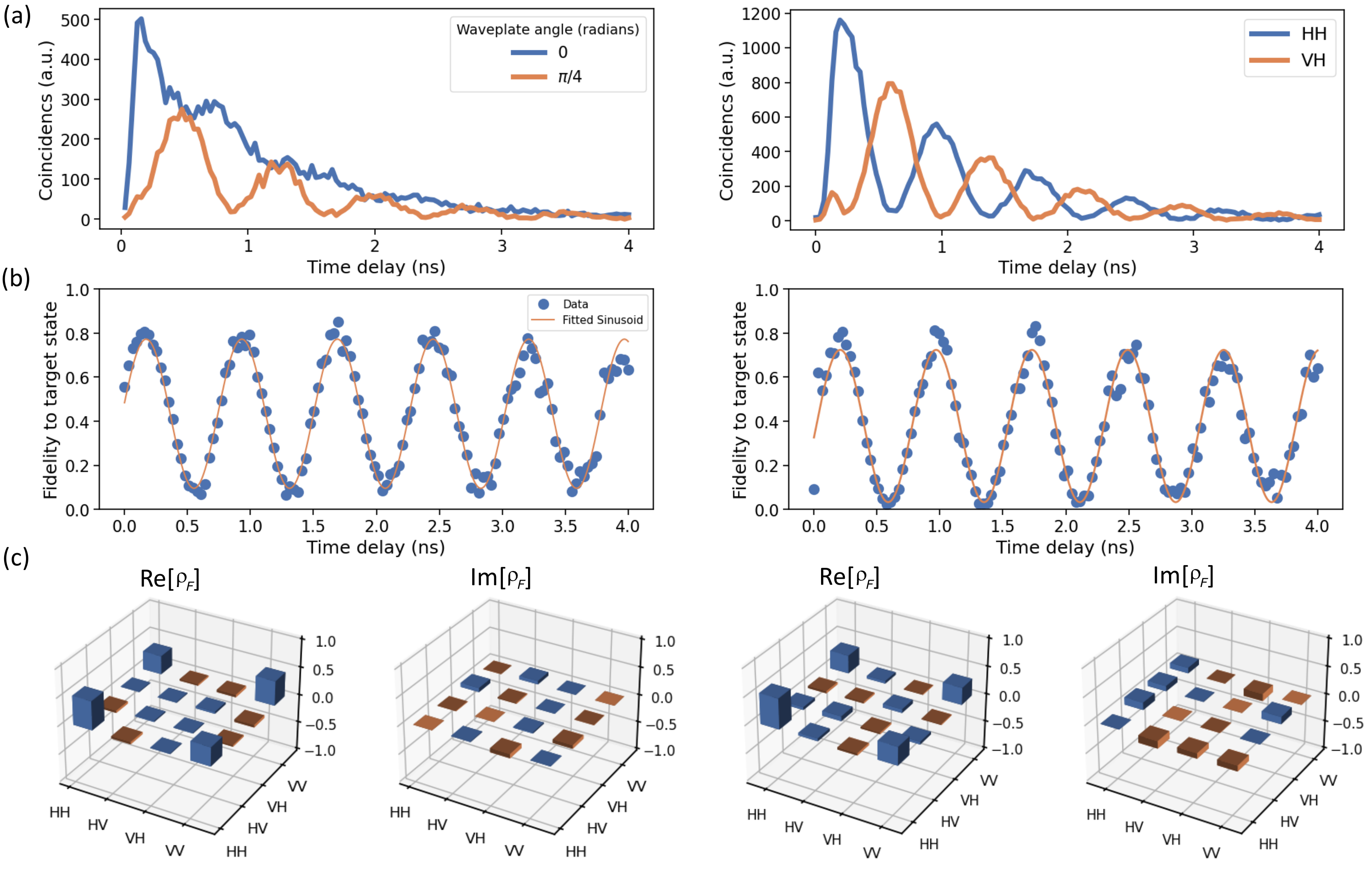}
\caption{\label{fig:QD_projection_fourier}
\textbf{Fourier (left) and projective (right) of the entangled pairs emitted by the quantum dot.
} (a) Coincidences measured at \num{0} and $\pi/4$ waveplate angles in the FT-QST (left), and at HH and VH projection bases (right). (b) Fidelity to the $\Phi^+$ Bell state calculated from the reconstructed density matrices in both methods at each time bin across the cascade. For the time delay of $\sim$ \SI{208}{\pico\second}, we obtain a fidelity of $0.807 \pm 0.0002$ from Fourier based reconstruction, and $0.806 \pm 0.0017$ from projective reconstruction. Both measurements accurately capture the FSS induced time evolution of the fidelity. A sinusoidal fit reveals a FSS of \num{5.44} $\mu$eV and \num{5.43} $\mu$eV from FT-QST and projective tomography, respectively.
}
\end{figure*} 
In Figure \ref{fig:single_photon_Polarimetry}, we present typical measurement for the single-photon polarization states $\ket{H}$ and $\ket{L}$. The incident photon rate for this measurement was set to \num{30} kcps, with an integration time of \SI{30}{\milli\second} per data point. The QWP rotated from \num{0} to $\pi$, and the signal was sampled at 400 evenly spaced points over this range, corresponding to a total measurement time of 20 seconds per polarization state. Figure \ref{fig:single_photon_Polarimetry}.a, presents the signal measured by the SNSPD for $\ket{H}$ (top) and $\ket{L}$ (bottom) polarization states as the waveplate rotated from $0$ to $\pi$. The measured signal, shown in blue dots, agrees well with the simulated expectation represented by the orange line for both states. The real and imaginary parts of the reconstructed density matrices from Fourier analysis are shown in \ref{fig:single_photon_Polarimetry}.b for $\ket{H}$ (top) and $\ket{L}$ (bottom) polarizations. For the $\ket{H}$ measurement, we calculated a fidelity of \num{0.999} to the ideal $\ket{H}$ state, while for $\ket{L}$, we calculated a fidelity of \num{0.997} from the Fourier reconstruction, confirming the high accuracy of the measurement.
The SNSPD, with a time resolution of approximately \SI{30}{\pico\second} and a detection efficiency of 25$\,\%$ at a dark count rate of \num{30} s$^{-1}$, enabled precise measurements even at low photon rates. To further assess the robustness of the FT-QST method and its performance at low light intensities, we performed a series of measurements varying the incident photon rate. We began with an incident photon rate of \num{1000} cps, for which we obtained a fidelity of \num{0.995} to the ideal $\ket{H}$ state with an integration time of \SI{30}{\milli\second} per point and a total measurement time of \SI{12}{seconds}. We observe near unity fidelity at all light levels ranging from \num{1} to \num{150} kcps, confirming the accuracy of FT-QST even at significantly lower intensities. Furthermore, we repeated the measurement for a $\ket{V}$ polarization state \num{200} times as shown in Figure \ref{fig:single_photon_Polarimetry}.c. The results demonstrate consistent readout of the state, with a negligible standard deviation of $0.0004$ from an average fidelity of $0.9964$, confirming the stability and reliability of FT-QST. 

\section*{Two-photon tomography}
As mentioned earlier, FT-QST is particularly advantageous when used to measured multi-photon states. Here we present results from two-photon state tomography experiments performed using entangled pairs emitted by an InAs/GaAs semiconductor quantum dot and a SPDC source, both emitting in the telcom C-band. For benchmarking, both standard FT-QST and projective tomography were used to characterize the entangled states.

\subsection*{Quantum dot source}
Quantum dots stand out as a promising platform for generating polarization-entangled photon pairs. This is achieved through the decay of a biexciton (XX) cascade to the ground state via an intermediate exciton (X) level \cite{Zeuner2021Aug}]. Ideally, this process produces the Bell state $\ket{\Phi^+}=\frac{1}{2}(\ket{HH}+\ket{VV})$. However, asymmetries in the quantum dot confinement potential generally lift the exciton level degeneracy, introducing a fine-structure splitting (FSS) that induces a time-dependent relative phase, $\phi(t)=tFSS/\hbar$, on the entangled state, leading to the time-evolving entangled state $(\ket{\Phi^+}=\frac{1}{2}(\ket{HH}+e^{i\phi(t)}\ket{VV})$, where $t$ represents the time delay between XX and X decay \cite{Ward2014Feb, Zeuner2021Aug}. By using high-speed SNSPDs, this phase  evolution can be resolved \cite{Fognini2019Jul, Alqedra2025Feb}.
\begin{figure*}[htb]
\includegraphics[width=2\columnwidth]{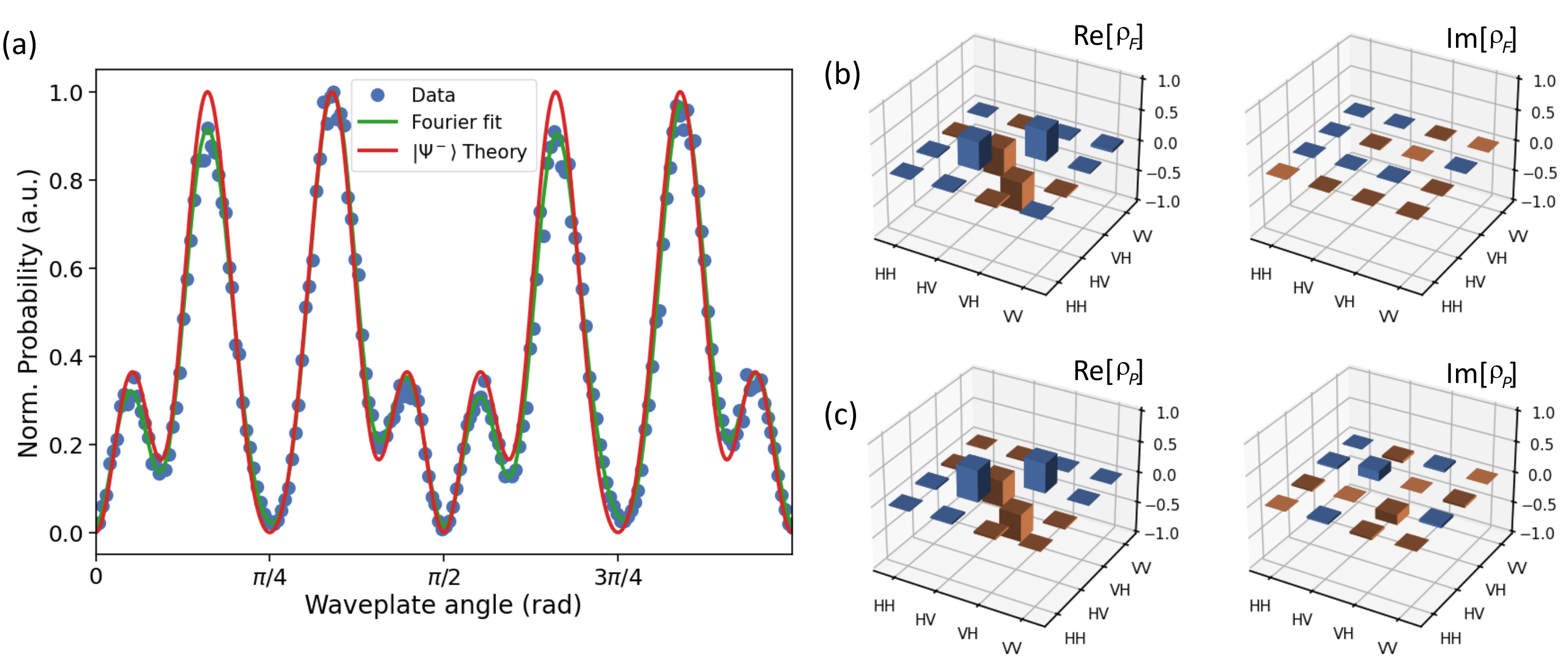}
\caption{\label{fig:spdc_projection_fourier}
\textbf{Fourier and projective tomography of entangled pairs generated by an SPDC source.} (a) Measured (blue dots), Fourier fit (green line) of the SPDC source as well as the theoretical values (red line) for the ideal Bell state $\ket{\Psi^-}$. (b) Real and imaginary parts of the reconstructed density matrix from Fourier and (c) projective tomography. Positive values are shown in blue, while negative values are represented in orange.
}
\end{figure*} 
For FT-QST, we used two quarter waveplates: the slower waveplate rotated from \num{0} to $\pi$, while the faster waveplate rotated five times that range, from \num{0} to $5\pi$. Photon coincidences were measured at \num{100} evenly spaced angles of the slower waveplate, with an integration time of \SI{25}{minutes} per point, enabling the reconstruction of the quantum state through Fourier analysis. We performed Fourier analysis on the coincidences and used the maximum-likelihood estimation algorithm to reconstruct the physical density matrix that best described the state. For projective tomography, we measured a total of 16 projection bases, with an integration time of 1 hour per base, and reconstructed the density matrices at each time bin across the cascade emission using the maximum likelihood estimation algorithm in Ref. \cite{afognini2025Jan}. To account for systematic polarization rotations and birefringence in our setup, we applied a virtual waveplate correction to the raw density matrices at all time bins for both FT-QST and projective tomography measurements. For FT-QST, the correction angles were $\theta=0.507$ rad and $\phi=-0.270$ rad , while for projective tomography, the angles were 
$\theta=-0.08$ rad and $\phi=1$ rad. This correction ensures that the reconstructed states are accurately aligned with the intended polarization bases, minimizing systematic errors in the measurements \cite{afognini2025Jan, Zeuner2021Aug}.

Figure \ref{fig:QD_projection_fourier} presents the result from FT-QST (left panel) and projective tomography (right panel).
Figure \ref{fig:QD_projection_fourier}.a, displays the coincidences measured at \num{0}, $\pi/4$ waveplate angles from FT QST (left), and the $\ket{HH}$ and $\ket{VH}$ projection basis from the projective tomography (right). The observed modulation in both cases arises from the time-varying phase due to the FSS. 
The fidelity calculated from the reconstructed density matrices at each time bin across the cascade for both tomography methods is shown in Figure \ref{fig:QD_projection_fourier}.b. The fidelity curves were fitted to a sinusoidal function, yielding a FSS 5.44 $\mu$eV from FT QST, and 5.43 $\mu$eV from projective measurement. The consistency in FSS estimates underscores the robustness of FT-QST in capturing the temporal evolution of the quantum state. Furthermore, within each time bin, the two methods exhibit excellent agreement between the calculated fidelity, well within the error bars. Finally, Figure \ref{fig:QD_projection_fourier}.c displays the real and imaginary parts of the density matrix reconstructed using both methods at a time delay of \SI{208}{\pico\second}. Positive values are shown in blue, while negative values are represented in orange. We obtain a fidelity of $0.807 \pm 0.0002$ to the Bell state $\Phi^+$ and a concurrence of $0.64 \pm 0.0011$ from Fourier based reconstruction, and a fidelity of $0.806 \pm 0.0017$ and a concurrence of $0.686 \pm 0.025$ from projective reconstruction.  

\subsection*{Spontaneous parametric down-conversion source}

To further demonstrate the versatility of FT-QST, we applied the technique to entangled photon pairs generated by a SPDC source in the telecom C-band \cite{Kwiat1995Dec, Fedrizzi2007Nov}. 
The quantum source is based on a periodically poled potassium titanyl phosphate (PPKTP) crystal pumped at 785 nm. The PPKTP crystal is placed in a Sagnac interferometer, which enables counter-propagating pumping to generate polarization-entangled photon pairs at telecom wavelengths through a type-II SPDC process. Unlike the time-evolving states produced by quantum dots, SPDC sources generate stationary entangled states. The target state of the SPDC source used here is the Bell state $\ket{\Psi^-} = \frac{1}{\sqrt{2}}(\ket{HV} - \ket{VH})$.
Figure \ref{fig:spdc_projection_fourier}.a shows the measured coincidence counts as a function of waveplate angle, rotated from \num{0} to ($\pi$) in \num{100} evenly spaced steps, with an integration time of half a second per point. The fitted curve derived from Fourier-transform reconstruction (green line), closely aligns with the expectation for the Bell state $\ket{\Psi^-}$ (red line), indicating high fidelity to the target state $\ket{\Psi^-}$.

To quantify the entanglement and fidelity of the SPDC-generated state, we reconstructed the density matrix using FT-QST (Figure. \ref{fig:spdc_projection_fourier}.b). The matrix exhibits the characteristic antisymmetric coherence terms and zero population in the $\ket{HH}$ and $\ket{VV}$ components, consistent with the $\ket{\Psi^-}$ Bell state. From the reconstructed density matrix, we calculated a fidelity of $0.941 \pm 0.006$ to the target state $\ket{\Psi^-}$ and a concurrence of $0.904 \pm 0.018$. 

To further validate the FT-QST results, we performed projective state tomography by measuring \num{16} projection bases, with an integration time of hald a second per projection. The reconstructed density matrix from this measurement is shown in Figure \ref{fig:spdc_projection_fourier}.c. We obtained a fidelity of $0.925 \pm 0.001$, and a concurrence of $0.923 \pm 0.011$, in a close agreement with those obtained from FT-QST. The slight discrepancies between the two methods are attributed to experimental variations, such as changes in fiber coupling efficiency and polarization alignment introduced by the apparatus. In particular, the insertion of the FT-QST apparatus into the SPDC setup can slightly alter the fiber coupling, thereby affecting the measured state.

In conclusion, we demonstrated the first experimental realization of Fourier-transform quantum state tomography for single- and two-qubit states \cite{Mohammadi2013Jan}. By measuring a pseudocontinuous time-dependent signal as waveplates rotate, FT-QST significantly reduces experimental overhead compared to conventional projective tomography, scaling linearly with the number of qubits rather than exponentially. We validated the technique using polarized weak coherent states and polarization entangled pairs generated by both quantum dots and SPDC sources. The reconstructed density matrices from FT-QST show excellent agreement with those obtained through projective tomography. 
The robustness of FT-QST across different types of entangled states highlights its versatility, accuracy and reliability. Future extensions of FT-QST could explore its application to higher-dimensional systems, such as qutrits and qudits. These results establish FT-QST as a resource-efficient and scalable alternative for characterizing multi-qubit systems, with significant potential for applications in quantum communication, cryptography, and photonic quantum computing. 

\begin{acknowledgments}
    A.W.E acknowledges support from Knut and Alice Wallenberg (KAW) Foundation through the Wallenberg Centre for Quantum Technology (WACQT).
\end{acknowledgments}

\bibliography{met_sync}
\end{document}


\title{Fourier Transform Quantum State Tomography of Polarization-Encoded Qubits}
\author{Mohammed K. Alqedra}
 \thanks{These authors contributed equally to this work.}
 \affiliation{Department of Applied Physics, Royal Institute of Technology, Albanova University Centre, 10691, Stockholm, Sweden}

\author{Pierre Brosseau}
 \thanks{These authors contributed equally to this work.}
 \affiliation{Centre for Disruptive Photonic Technologies, TPI, Nanyang Technological University, Singapore}
 \affiliation{Division of Physics and Applied Physics, SPMS, Nanyang Technological University, Singapore}

\author{Ali W. Elshaari}
 \affiliation{Department of Applied Physics, Royal Institute of Technology, Albanova University Centre, 10691, Stockholm, Sweden}

\author{Jun Gao}
 \affiliation{Department of Applied Physics, Royal Institute of Technology, Albanova University Centre, 10691, Stockholm, Sweden}

\author{Anton N. Vetlugin}
 \affiliation{Centre for Disruptive Photonic Technologies, TPI, Nanyang Technological University, Singapore}
 \affiliation{Division of Physics and Applied Physics, SPMS, Nanyang Technological University, Singapore}

\author{Cesare Soci}
 \affiliation{Centre for Disruptive Photonic Technologies, TPI, Nanyang Technological University, Singapore}
 \affiliation{Division of Physics and Applied Physics, SPMS, Nanyang Technological University, Singapore}

\author{Val Zwiller}
 \email{zwiller@kth.se}
 \affiliation{Department of Applied Physics, Royal Institute of Technology, Albanova University Centre, 10691, Stockholm, Sweden}


\date{\today}
\maketitle

\section{Theoretical framework for Fourier state tomography}
Following Mohammadi \emph{et al.}~\cite{Mohammadi2013Jan}, the time-dependent function, $\chi_{m,i_m}(t)$, that encodes the effect of the rotating waveplate in each mode can be written as:
\begin{equation}
    \chi_{m,i}(t) = \text{Tr} \big[ \sigma_i M_m^H (t) \big].
\end{equation}

Here, the measurement operator $M_m^H (t)$ is defined as:
\begin{equation}
    M_m^H (t) = U_m^\dagger (t) |H\rangle \langle H | U_m (t),
\end{equation}

where $U_m (t)$ is the time-dependent unitary transformation associated with the rotating waveplate: 
\begin{equation}
    \hat{U}_m(t) \;=\;
    \cos\!\Bigl(\frac{\beta}{2}\Bigr)\,\hat{\sigma}_0
    \;-\; i\,\sin\!\Bigl(\frac{\beta}{2}\Bigr)\,\bigl[\vec{v}_m(t)\cdot\vec{\sigma}\bigr].
    \label{eq:Um_operator}
\end{equation}

Here, $\hat{\sigma}_0$ is the identity operator, $\vec{\sigma} = (\sigma_x,\sigma_y,\sigma_z)$ is the vector of Pauli matrices. The parameter $\beta$ represents the retardance of the waveplate, which determines the phase shift introduced between orthogonal polarization components as the waveplate rotates. The vector
\begin{equation}
    \vec{v}_m(t)
    = \cos(\omega_m t)\,\hat{k}
      \;+\;\sin(\omega_m t)\,\hat{i}
    \label{eq:rotation_vector}
\end{equation}

defines the instantaneous rotation axis in the Bloch-sphere picture. Here,
$\hat{k}$ and $\hat{i}$ are orthogonal unit vectors in Euclidean space.

\subsection*{Single-qubit states}
For single-qubit states, this Fourier tomography process is mathematically equivalent to polarimetry, where the Stokes parameters describe the polarization state of light. We can express the intensity at the detector as a Fourier series whose coefficients can be related to the polarization state of light.
\begin{align}
P_{\text{single}} &= \langle \Psi | M(\theta) | \Psi \rangle \nonumber \\
&= A_0 + A_{4\theta} \cos(4\theta) + B_{2\theta} \sin(2\theta) \nonumber \\
&\quad + B_{4\theta} \sin(4\theta)
\end{align}

By simple inversion, we can express the density matrix entries of the state being measured as functions of the Fourier coefficients of the series.
\[
p = \begin{bmatrix}
p_{11} & p_{12} \\
p_{21} & p_{22}
\end{bmatrix}
\]
with:
\begin{align}
p_{11} &= A_0 + A_{4\theta}, \\
p_{12} &= \frac{B_{2\theta} + 2i B_{4\theta}}{i}, \\
p_{21} &= \frac{B_{2\theta} - 2i B_{4\theta}}{-i}, \\
p_{22} &= A_0 - 3 A_{4\theta}
\end{align}

The coefficients \( A_0 \), \( A_{4\theta} \), \( B_{2\theta} \), and \( B_{4\theta} \) are obtained by Fourier analysis of the detected signal.

\subsection*{Two-qubit states}
This approach is also valid for multi-photon states. Specifically, in the two-qubit case, we can express the coincidences between the two detectors as a Fourier series whose coefficients can be related to the two-qubit polarization state:
\begin{align}
P_{\text{single}} &= \langle \Psi | M(\theta) | \Psi \rangle \nonumber \\
&= A_0 + \sum_{\substack{n=2 \\ n \neq 5}}^{12} A_{2n\theta} \cos(2n\theta) + \sum_{\substack{n=1 \\ n \neq 4,6}}^{12} B_{2n\theta} \sin(2n\theta).
\end{align}

In the same way, we can express each entry of the density matrix as functions of the Fourier coefficients of the series.
\begin{align}
p &= \begin{bmatrix}
p_{11} & p_{12} & p_{13} & p_{14} \\
p_{21} & p_{22} & p_{23} & p_{24} \\
p_{31} & p_{32} & p_{33} & p_{34} \\
p_{41} & p_{42} & p_{43} & p_{44}
\end{bmatrix}
\end{align}

\begin{align}
p_{11} &= a_0 + a_4 + a_{20} + a_{16} + a_{24}, \\
p_{12} &= -i b_{10} - 2i b_{14} + 2 b_{20} + 2 b_{16} + 2 b_{24}, \\
p_{13} &= -2 b_{16} + 2i b_{18} - i b_2 + 2 b_{24} + 2 b_4, \\
p_{14} &= \frac{1}{4} \left( 8 a_{12} + 16 a_{16} - 16i a_{18} - 16 a_{24} - 16i a_6 \right), \\
p_{21} &= i b_{10} + 2i b_{14} + 2 b_{20} + 2 b_{16} + 2 b_{24}, \\
p_{22} &= a_0 + a_4 - 3 a_{20} - 3 a_{16} - 3 a_{24}, \\
p_{23} &= \frac{1}{4} \left( -8 a_{12} + 16 a_{16} - 16i a_{18} - 16 a_{24} + 16i a_6 \right), \\
p_{24} &= 6 b_{16} - 6i b_{18} - i b_2 - 6 b_{24} + 2 b_4, \\
p_{31} &= -2 b_{16} - 2i b_{18} + i b_2 + 2 b_{24} + 2 b_4, \\
p_{32} &= \frac{1}{4} \left( -8 a_{12} + 16 a_{16} + 16i a_{18} - 16 a_{24} - 16i a_6 \right), \\
p_{33} &= a_0 - 3 a_{16} + a_{20} - 3 a_{24} - 3 a_4, \\
p_{34} &= -i b_{10} + 6i b_{14} + 2 b_{20} - 6 b_{16} - 6 b_{24}, \\
p_{41} &= \frac{1}{4} \left( 8 a_{12} + 16 a_{16} + 16i a_{18} - 16 a_{24} + 16i a_6 \right), \\
p_{42} &= 6 b_{16} + 6i b_{18} + i b_2 - 6 b_{24} + 2 b_4, \\
p_{43} &= i b_{10} - 6i b_{14} + 2 b_{20} - 6 b_{16} - 6 b_{24}, \\
p_{44} &= a_0 + 9 a_{16} - 3 a_{20} + 9 a_{24} - 3 a_4.
\end{align}

The coefficients \( A_i \) and \( B_i \) are obtained through Fourier analysis of the coincidence signal.

\section{Quantum dot characterization setup}

\begin{figure}[htb]
\centering
\includegraphics[width=\columnwidth]{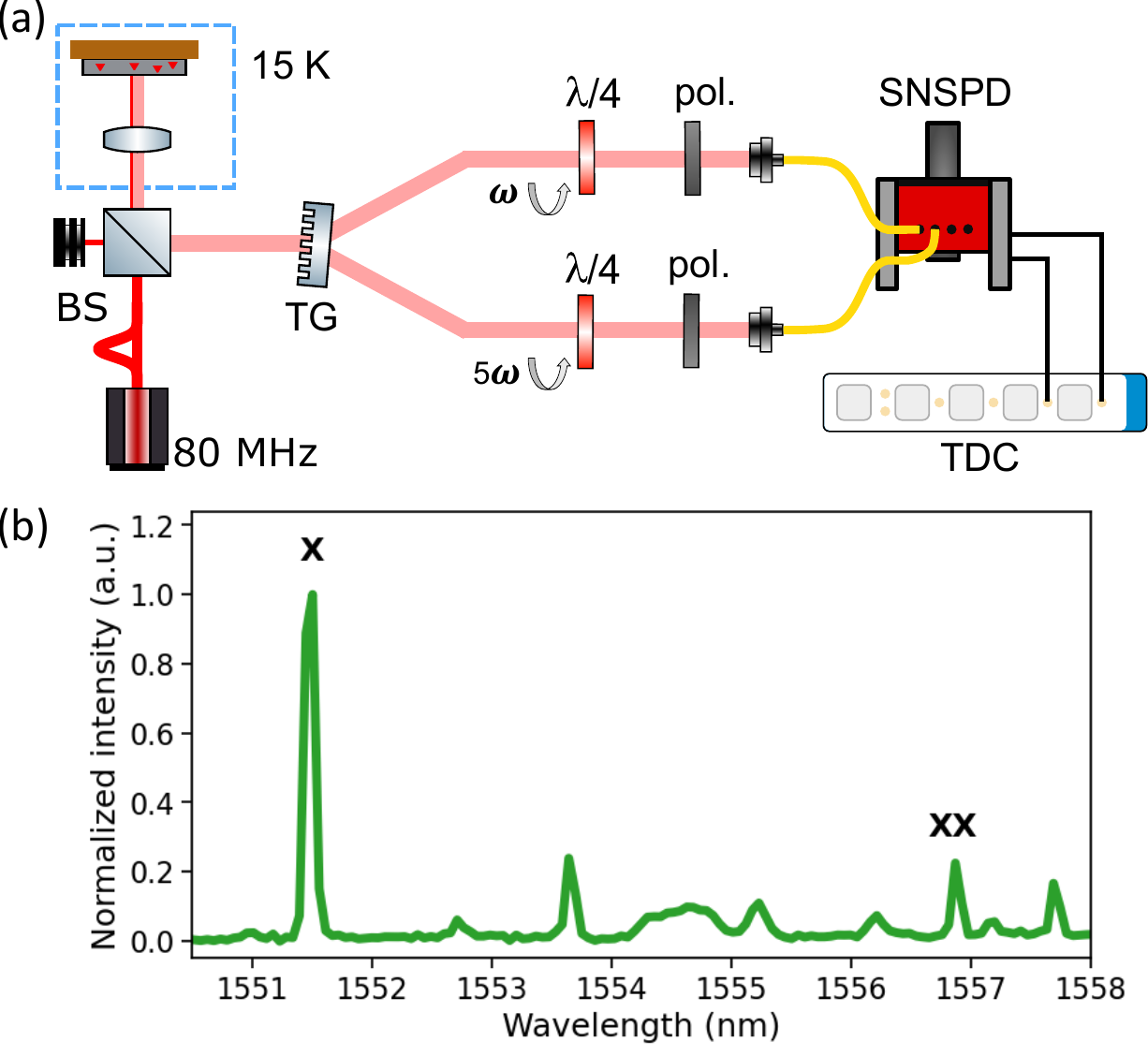}
\caption{\label{fig:QD_setup}
(a) Schematic of the setup used to characterize the quantum dot. (b) The QD emission spectrum under p-shell excitation measured with an InGaAs spectrometer.
}
\end{figure}

The setup is designed to excite the quantum dots (QDs) and collect their emission through a confocal microscopy configuration, as shown in Figure \ref{fig:QD_setup}(a). An 80 MHz picosecond pulsed laser source is used for excitation. The laser beam passes through a 10:90 beam splitter (BS), where 10$\,\%$ of the light is directed toward the QD sample, and 90$\,\%$ of the QD emission is collected for measurements. The QDs are excited to the p-shell at a wavelength of 1470 nm, and the emission is collected using a high numerical aperture (NA = 0.8) objective. After collection, the exciton (X) and biexciton (XX) emission lines are spectrally separated using a transmission grating (TG). The separated emission lines are then directed toward the Fourier-transform quantum state tomography (FT-QST) setup. The FT-QST setup consists of two rotating quarter-wave plates (QWPs), with one waveplate rotating at a frequency $\omega$ and the other at $5\omega$, followed by a fixed polarizer. The waveplates are rotated in 100 discrete steps, and the photons are detected using SNSPDs. The time-resolved measurements are recorded using a time-to-digital converter (qutools quTAG), and the data are analyzed using the Extensible Timetag Analyzer ETA software \cite{Lin2021Aug}.
\\
\bibliography{met_sync}